# Quantifying Locomotion Differences Between Virtual Reality Users With and Without Motor Impairments

Rachel L. Franz

Computational Media and Arts | Internet of Things, Hong Kong University of Science and Technology (Guangzhou), Guangzhou, Guangdong, China, rachelfranz@hkust-gz.edu.cn

Jacob O. Wobbrock

The Information School | DUB Group, University of Washington, Seattle, Washington, USA, wobbrock@uw.edu

Today's virtual reality (VR) systems and environments assume that users have typical abilities, which can make VR inaccessible to people with physical impairments. However, there is not yet an understanding of *how* inaccessible locomotion techniques are, and which interactions make them inaccessible. To this end, we conducted a study in which people with and without upper-body impairments navigated a virtual environment with six locomotion techniques to quantify performance differences among groups. We found that groups performed similarly with *Sliding Looking* on all performance measures, suggesting that this might be a good default locomotion technique for VR apps. To understand the nature of performance differences with the other techniques, we collected low-level interaction data from the controllers and headset and analyzed interaction differences with a set of movement-, button-, and target-related metrics. We found that movement-related metrics from headset data reveal differences among groups with all techniques, suggesting these are good metrics for identifying whether a user has an upper-body impairment. We also identify movement-, button, and target- related metrics that can explain performance differences between groups for particular locomotion techniques.

CCS CONCEPTS • **Human-centered computing~Accessibility~Empirical studies in accessibility** • **Human-centered computing ~ Human computer interaction (HCI)~Interaction paradigms~Virtual reality**

**Additional Keywords and Phrases:** Virtual Reality, Accessibility, Locomotion Techniques

## 1 INTRODUCTION[1]

Virtual Reality (VR) is no longer a niche technology and is rapidly expanding into mainstream use, with applications in gaming, education, healthcare, and beyond. The global VR user base is expected to grow by nearly 20% in the next four years [34], signaling a future in which immersive experiences are part of everyday life. Yet as VR adoption accelerates, so do concerns about who gets to participate.

Locomotion, which refers to how users move through virtual environments, is one of the most fundamental and frequent interactions in VR. A wide array of locomotion techniques has emerged from years of research and development [38]. However, these techniques often presume users have typical motor abilities, making them difficult or impossible to use for people with physical impairments, particularly upper-body motor impairments. Since most VR systems rely on handheld controllers and head-tracked input, people with limited range of motion, muscle weakness, or coordination difficulties may encounter significant challenges using these systems.

---

[1] Portions of this paper were published as Franz, R. L. (2024*). Supporting the Design, Selection, and Evaluation of Accessible Interaction Techniques for Virtual Reality*. (Publication No. 31490607) [Doctoral dissertation, University of Washington]. University of Washington ProQuest Dissertations & Theses. This work also builds upon prior work: Rachel L. Franz, Jinghan Yu, and Jacob O. Wobbrock. 2023. Comparing locomotion techniques in virtual reality for people with upper-body motor impairments. In Proceedings of the Conference on Computers and Accessibility (ASSETS'23), 1–15. https://doi.org/10.1145/3597638.3608394

Although prior research has identified a range of accessibility challenges [16, 43, 58], we still lack a clear understanding of *how* inaccessible VR locomotion techniques actually are. Establishing a quantitative baseline comparison between users with and without upper-body impairments is a critical step toward identifying which techniques are currently exclusionary and which might offer a more equitable experience. If we can identify techniques where users with and without impairments perform similarly, we can begin to define what inclusive locomotion looks like in practice.

Beyond measuring *if* performance differs, it is equally important to understand *why*. Low-level interaction data, such as headset motion and controller input, can reveal how impairments affect the way people use different locomotion techniques. For example, reduced head rotation during a head-based technique could suggest neck mobility limitations that impact performance. By analyzing these data, we can uncover specific mismatches between user abilities and system demands.

Upper-body motor impairments vary widely in type and severity, but certain patterns of difficulty may emerge across different conditions. Identifying interaction-level metrics that consistently signal accessibility issues regardless of specific diagnoses, could help designers pinpoint which aspects of a locomotion technique need adjustment. While most previous work has focused on qualitative accounts of VR accessibility [16, 43], our approach brings a quantitative lens to understanding accessibility in VR locomotion.

Finally, these metrics have the potential to go beyond analysis and into prediction. If interaction-level metrics can reliably indicate likely performance with a given technique, they could power adaptive systems that recommend or adjust locomotion techniques in real time. This opens the door to VR systems that don't just accommodate impairments, but also respond to them, creating more inclusive virtual experiences for all.
Based on these points, we present the following research questions.

(RQ.1) *How do people with and without upper-body impairments perform across six representative locomotion techniques?*

(RQ.2) *Does any single locomotion technique enable comparable performance for users regardless of ability, pointing to a universally inclusive solution?*

(RQ.3) *Which low-level headset and controller metrics (e.g., device distance, velocity, acceleration, etc.) best explain the observed performance differences between groups, and how can these insights guide accessible locomotion technique designs?*

We conducted a study comparing six locomotion techniques in which we recorded high-level performance measures (e.g., trial completion time, target hit rate, and obstacle hit rate) as well as low-level movement-, button-, and target-related metrics with data from the headset and controllers of 20 people with and 20 people without upper-body impairments ($N = 40$). We compared these measures between groups and found that with one of the techniques, *Sliding Looking*, participants performed similarly on all performance measures. We also found that some metrics revealed *why* the two groups performed differently with the other five locomotion techniques.

We make three contributions in this work: (1) We report results from a study comparing performance differences of people with and without upper-body impairments when they used six VR locomotion techniques. (2) We also contribute a set of movement-, button-, and target-related metrics that can be used to measure interaction differences between people with and without impairments when they use locomotion techniques. (3) Finally, we report results comparing how people interacted with their headset and controllers with a set of movement-, button-, and target-related metrics to explain performance differences with the six locomotion techniques.

## 2 RELATED WORK

In the past few years, researchers have advanced an understanding and improvement of the accessibility of VR for people with a range of disabilities, particularly concerning vision [3–6, 18, 46, 62, 63] and hearing [2, 30, 35–37, 41]. This section focuses on the work related to VR accessibility for people with physical impairments as well as how movement is measured when people interact with touchscreen and desktop devices.

### 2.1 Accessibility in Virtual Reality for People with Physical Impairments

As VR continues to gain popularity, researchers have prioritized making VR accessible to users with disabilities, including to those with physical impairments. Yet, even with notable advancements, current VR hardware and interaction techniques still assume users can perform actions like standing, walking, reaching, and crouching [16, 58]. As a result, people whose physical abilities differ from what the VR system expects can face significant accessibility challenges.

VR resembles other technologies in terms of audiovisual features, which suggests that designers should be able to adapt existing assistive tools designed for sensory accessibility for VR [63]. However, VR relies on three-dimensional interactions to create a convincing sense of realism, differentiating it from touchscreen or desktop paradigms. Consequently, research has been necessary to uncover physical accessibility barriers, particularly those impacting wheelchair users and individuals with upper-body impairments.

### 2.2 2.1.2 Identifying Accessibility Challenges

Research highlights numerous accessibility barriers encountered by individuals with limited physical mobility when using commercial VR systems. Key difficulties include wearing and removing headsets, pressing controller buttons, and keeping track of virtual controllers [43]. Additionally, findings from the Disability Visibility Project [58] reported accessibility barriers arising from assumptions embedded in motion-tracking systems, such as requiring users to stand or use both hands. Wheelchair users faced particular difficulties performing movements that the VR system recognized. This accessibility problem was compounded by software restrictions in commercial VR systems preventing the use of alternative input devices like gamepads. Participants have also reported that VR systems often fail to accommodate varying degrees of dexterity, reflexes, and range of motion [12, 58].

Gerling and Spiel [16] further analyzed these issues, identifying ableist design assumptions that overlooked diverse physical capabilities. They proposed that designers explicitly examine "actions lent", or the physical actions necessary for interacting with VR, to uncover hidden barriers and promote inclusive design practices.

### 2.3 2.1.3 Enhancing Accessibility

Several solutions have emerged to address accessibility challenges for wheelchair users. WalkinVR, a commercial tool, enables users to customize mappings between their physical movements and virtual controller actions. Gerling et al. [15] created VR game prototypes explicitly designed for wheelchair users and stressed the need for adaptable interfaces to accommodate different impairments. Additionally, Sassi et al. [47] studied redirected wheelchair movements and found that participants did not notice redirection if it was subtle, indicating promise for the use of redirection to make locomotion accessible.

Another frequent obstacle for wheelchair users is their lower vantage point, which makes interactions that require a higher viewpoint challenging. Taheri et al. [52] documented positive experiences when wheelchair users virtually experienced walking, noting benefits from elevated perspectives. Qorbani et al. [45] and Ganapathi et al. [14] enabled seated users to adjust their virtual height, which significantly enhanced their ability to perform tasks such as object retrieval. Researchers have also explored alternative techniques for individuals with lower-body impairments who do not use wheelchairs. Mahmud et al. [40] showed that people with multiple sclerosis increased their walking speed in VR when they were provided with spatial audio feedback.

Efforts to improve accessibility for individuals with upper-body impairments have included frameworks that accommodate single-handed interactions. For instance, Yamagami et al. [60] introduced guidelines for translating two-handed into single-handed tasks. Other frameworks have emerged to help designers select and create scene-viewing techniques that do not rely on head movements [10, 11]. Researchers have also asked participants with upper-body impairments to design gestures and found that people created various head, mouth, and hand gestures [53, 61]. Finally, a study comparing the accessibility of six locomotion techniques with people with upper-body impairments found that participants were faster and more accurate when navigating to targets when they used locomotion techniques that required button input and minimal movement [12].

**2.4  Measuring the Movement Characteristics of People With and Without Physical Impairments**

Understanding how people with motor impairments move in 2-D and 3-D space has been a goal for many researchers. Characterizing movement is achieved through the design of metrics that capture differences between individuals with various types of impairments and people without impairments.

Mackenzie et al. [39] proposed a set of seven accuracy measures to explain performance differences in terms of speed and accuracy of input devices for people without impairments. They found that two measures, "movement offset" (how far the user's path deviated from the optimal path) and "target re-entry" (how many times a participant exited then entered a target) were correlated with speed and accuracy. Based on this study, Keates et al. [31] proposed a set of measures to characterize the movement patterns of people with motor impairments. They found that path metrics such as the distance travelled relative to the cursor displacement and the distribution of distance travelled for a range of cursor speeds could reveal differences between participants with and without motor impairments. These measures could also encode differences between participants with a greater impairment severity compared to those who have a less severe impairment [31].

Hurst et al. [28] predicted whether people with motor impairments would need an accessibility adaptation when using a mouse. They built machine learning models that achieved 83% accuracy when classifying whether a user could benefit from Steady Clicks, an accessibility feature that keeps the cursor stationary while the user is clicking, preventing it from slipping off the target. They used feature selection to identify the best features for building a model and found that click-related features and task completion time were most predictive. They also built models to distinguish people with and without motor impairments and found that the task completion time and number of clicks, as well as target re-entry, were the most important features when predicting if a user had an impairment.

Instead of overall movement characteristics of mouse use, Hwang et al. [29] investigated how submovements, or movements between velocity dips, could reveal differences between people with and without motor impairments. They proposed a set of accuracy measures including "number of pauses" and "peak velocity of submovements," which provided insight into differences based on impairment type as well as differences between people with and without impairments.

Aside from cursor-related measures, touchscreen-related measures have also been proposed. A study by Kong et al. [33] proposed 15 touch metrics to characterize touches. They found that five metrics differentiated people with and without impairments including "touch variability," or the total distance the finger travelled during a touch, and "touch drift," or the Euclidean distance between the touch down and touch lift, among other measures [33].

While the above work compares movement in two dimensions, some researchers have also investigated movement characteristics in three dimensions. A study by Viau et al. [54] found that people with motor impairments were significantly slower when placing a ball on a virtual target compared to people without motor impairments. Also, the path their hand took was more curved compared to that of people without impairments. However, the authors noted that overall, the movement strategies for both groups were similar [54].

Our current study builds on prior work by investigating the physical accessibility of VR, particularly locomotion techniques. We also adapt measures from prior work on desktop and touchscreen devices to VR.

## 3 STUDY METHOD

This section methodologically replicates the procedure of a prior study [12] that evaluated six locomotion techniques with participants who have upper-body impairments. In the present study, we extend that work by comparing this same group to a new group of 20 participants without impairments. In this comparative analysis, we contribute a new evaluation of performance and perceived workload differences between groups to address RQ.1 and RQ.2. Furthermore, we introduce an in-depth quantitative analysis of low-level device data, such as headset and controller movements, to explain observed performance differences to address RQ.3. The study procedure was approved by the university ethics review board.

### 3.1 Participants

Twenty participants with upper-body impairments and 20 participants without upper-body impairments participated in the study ($N$ = 40). Sample size was estimated using G*Power[2] for an F test (repeated measures ANOVA: within-between interaction), with a medium effect size ($F$ = 0.25), two groups, six measurements, power set to 0.80, and alpha at 0.05. This analysis indicated that a sample of 20 participants would be sufficient to detect an effect of this size. As our study included 40 participants, we had twice the required sample size.

The upper-body impairments reported by participants affected various areas, including the neck, arms, and hands. All participants met general eligibility criteria, which included being at least 18 years old, fluent in English, and capable of providing informed consent. Additional criteria for the group with impairments required that participants have either a permanent or temporary upper-body motor impairment, be able to hold at least one Meta Quest 2 controller, and wear the Quest 2 headset.

The group with impairments consisted of 20 participants: four identified as non-binary, six as women, and ten as men. The average age in this group was 39.4 years ($SD$ = 16.6). Participants reported various conditions affecting motor function, which are detailed in Table 1. In terms of education, one participant had completed high school, three had some college experience, three held associate degrees, nine had bachelor's degrees, and four held master's degrees. When asked to rate their technology proficiency on a four-point scale (ranging from "not proficient" to "very proficient"), one participant identified as "not proficient," one as "somewhat proficient," six as "average," and eleven as "very proficient." One participant chose not to respond.

---
[2] https://www.psychologie.hhu.de/arbeitsgruppen/allgemeine-psychologie-und-arbeitspsychologie/gpower

Regarding prior experience with VR, ten participants had never used VR before. Six had used a headset, and four had used VR through both a headset and a phone. The year they first experienced VR ranged from 2016 to 2022. Frequency of use varied: three participants had tried VR only once, six used it occasionally, and one reported weekly usage.

Table 1. Participants' self-reported upper-body motor impairment conditions. From Franz et al. (2023).

| Participant | Condition |
|---|---|
| P00 | Left hand amputee |
| P01 | Spinal stenosis at 3C level |
| P02 | Ehler's Danlos Syndrome, Beals Syndrome (FBN2 gene mutation) causing progressive muscle weakness |
| P03 | C-5 quadriplegia |
| P04 | Peripheral neuropathy |
| P05 | Osteoarthritis, nerve damage |
| P06 | Hand tremor and weakness |
| P07 | Limb Girdle Muscular Dystrophy Type 2A |
| P08 | Cerebral palsy |
| P09 | Chronic joint pain, ongoing carpal tunnel syndrome |
| P10 | Paralysis, quadriplegia |
| P11 | Tetraplegia |
| P12 | Muscular dystrophy |
| P13 | Muscular dystrophy |
| P14 | C5-C6 incomplete spinal cord injury with functional quadriplegia |
| P15 | Nerve damage, severe muscle spasms, arthritis in neck, sacroiliac joint pain |
| P16 | Arthritis in both wrists, elbows, and shoulders from overuse, right arm fatigues easily |
| P17 | Peripheral neuropathy in all extremities |
| P18 | Transverse myelitis |

Within the group without impairments, 10 women and 10 men participated in the study. The mean age was 23.2 years ($SD$=5.0). Ten participants had a high school degree, two completed some college, four had a bachelor's degree, and four had a master's degree. On a four-point scale from "not proficient" to "very proficient" at using technology, one participant rated themselves as "somewhat proficient," 11 as "average," and eight as "very proficient."

Nine participants had never used VR, four participants had used VR with a headset, three had used VR with a smartphone, and four used VR with a headset and a phone. The first time they had ever used VR ranged from 2016 to 2022. Three participants reported using VR only once before, five reported using it rarely, one used it every few months, one used it one or more times a month, and one reported using it weekly.

### 3.2 Apparatus

Participants used a Meta Quest 2 to navigate a VE created with Unity. The environment comprised a virtual plane that contained six targets. Participants used each of six locomotion techniques to navigate the environment (Table 2). Locomotion techniques were selected based on a survey of existing techniques, with the goal of ensuring that participants engaged a range of body parts (e.g., head, arms, fingers) and motor abilities (e.g., gross and fine motor control) [38] (Figure 1).

Table 2. Locomotion techniques used in the study, their abbreviated names, and the instructions given to participants to control them. From Franz et al. (2023).

| Locomotion Technique | Instructions |
|---|---|
| Astral Body (Astral) | Use the thumbstick to move the avatar |
| Chicken Acceleration (Chicken) | Lean forward and look in the direction you want to move. Lean back to stop. |
| Grab and Pull (Grab) | Reach forward, press and hold the trigger, pull the controller back toward you, then release the trigger |
| Sliding Looking (Sliding) | Press and hold the "A" or "X" button and turn your head in the direction you want to move |
| Teleport (Teleport) | Aim the blue circle on the floor and press the trigger to jump to the circle |
| Throw Teleport (Throw) | Press and hold the grip to show a ball. Throw the ball and release the grip to release the ball. You will jump to where the ball lands. |

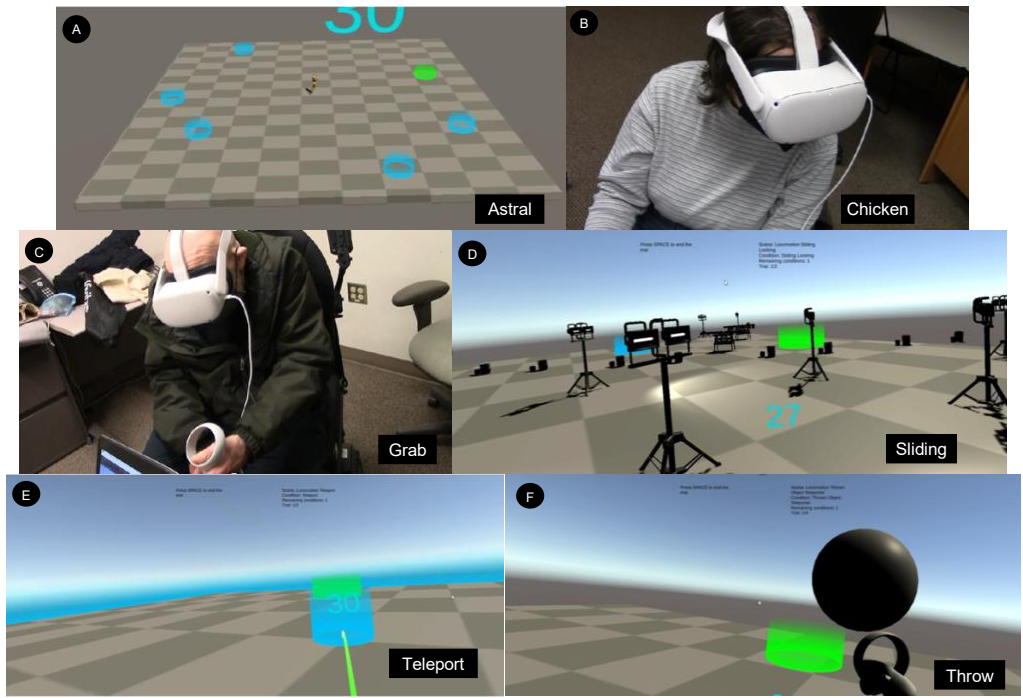

Figure 1. (A) The participant's perspective while using Astral, showing their avatar positioned at the center of the test environment, surrounded by targets. (B) The participant leaning forward and looking in the direction they want to move while using Chicken. (C) The participant reaching forward to grab the ground while using Grab. (D) The participant looking towards the target to maneuver around obstacles while using Sliding. (E) The participant aiming their teleporter toward the target when using Teleport. (F) The participant throwing the ball with their controller while using Throw.

### 3.3 Procedure

Participants provided written informed consent after being informed about the study's purpose, procedures, and their rights, including the option to withdraw at any time without penalty.

*Pre-task questionnaire*: Prior to using VR, two questionnaires were given to all 40 participants, both of which measured upper-body impairment level. The *Quick*DASH [19] was the first questionnaire participants responded to. It is validated and is a shortened version of the 30-item Disabilities of the Arm, Shoulder, and Hand (DASH)

Questionnaire [27]. The *Quick*DASH questionnaire was selected because it measures function in multiple areas of the upper body, is agnostic to medical diagnosis, and is short (only 11 items) [50]. Participants rated their responses on a Likert-type scale (1-5) ordered from low to high difficulty. An example item was, "Please rate your ability to do the following activities in the last week by circling the number next to the appropriate response: Use a knife to cut food." The full questionnaire is available in Appendix A1. An overall impairment score was calculated to represent upper-body function [27]. The score ranged between 0 to 100, where higher scores represented more functional impairment ($M$=27.1, $SD$=26.5).

The *Quick*DASH score was calculated for all 40 participants. Scores were then converted into four impairment levels to facilitate the interpretability of the data analysis. The four levels and corresponding *Quick*DASH scores were: low (0-25), medium-low (26-50), medium-high (51-75), and high (76-100). Due to the presence of only a single participant in the high group, this group was merged with the medium-high group to achieve a more balanced distribution of participants across groups. There were 21 participants in low, six in medium-low, and 10 in medium-high groups. The mean *Quick*DASH score for participants in the low group was 5.63 ($SD$=7.21), in the medium-low group was 41.48 ($SD$=7.66), and in the medium-high group was 61.18 ($SD$=10.54) (Figure 2).

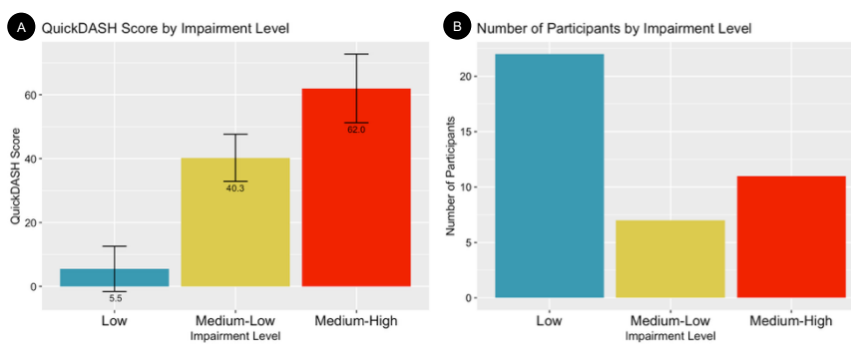

Figure 2. (A) Mean QuickDASH score by impairment level and (B) number of participants in each impairment level category.

The second questionnaire was a custom one based on literature in accessible computing. Specifically, it was a combination of participant characteristics reported by Mott et al. [43] and Findlater et al. [8] but adapted to questionnaire format. The impairments identified in the papers are likely to affect technology use, so we referred to it hereafter as the "**T**echnology-**R**elated physical **I**mpairment **Q**uestionnaire" (TRIQ). Participants indicated on 19 items whether they had a specific impairment, so each response was a dichotomous checkbox (i.e., checked or not). For example, some of the possible responses to the statement, "Check all the impairments you experience" were "slow movements," "tremor," and "poor coordination." The most common impairments were "low strength in core, shoulders, neck, arms, hands, or fingers" ($N$=19), "limited wrist extension or flexion" ($N$=18), and "difficulty holding objects" ($N$=18) (Figure 3). The average number of impairments participants selected of the 19 listed in the TRIQ was 5.8 ($SD$=6.3). The full questionnaire is available in Appendix A2. People with and without impairments filled out the *Quick*DASH and TRIQ before starting the study tasks. The *Quick*DASH took an average of 5.3 minutes ($SD$=2.9) and the TRIQ took an average of 6.9 minutes ($SD$=5.2) to complete.

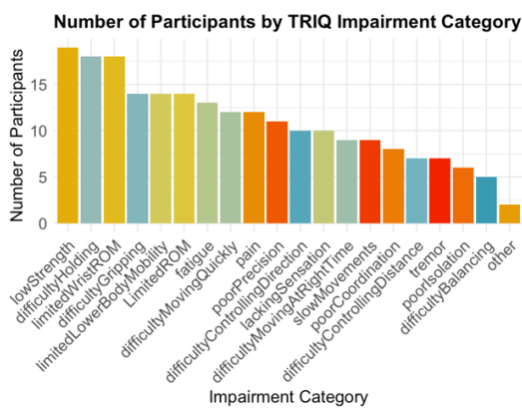

Figure 3. The number of participants that indicated they had a particular impairment category listed in the TRIQ.

*Locomotion task:* After completing the questionnaires, participants used the six locomotion techniques. The task they completed with each technique was to move to a target in the environment within 30 seconds and stay within the target for 5 seconds. They navigated to six different targets in the environment twice, for a total of 12 trials (Figure 4). In two of the trials, obstacles were presented, around which participants were instructed to navigate. After using each locomotion technique, they answered the post-task questionnaire.

*Post-task questionnaire*: The third questionnaire, referred to as the "post-task questionnaire," asked participants to rate their level of presence, simulator sickness, and workload they experienced using a locomotion technique [22, 32, 49]. The questionnaire included a single item to assess presence, adapted from question #1 of Slater and Steed's presence scale [48]. This specific item was chosen because prior research has shown it to be strongly discriminative [48]. Participants rated their feeling of presence on a scale from 1 ("not at all") to 7 ("very much"). To evaluate general simulator sickness, we also included the first item from the Simulator Sickness Questionnaire (SSQ) [32], which asks about overall discomfort, including nausea, dizziness, eyestrain, and vertigo. Responses were recorded on a 0 to 3 scale, with 0 indicating "no discomfort" and 3 indicating "severe discomfort." We included five items from the NASA-TLX workload assessment [21], omitting the temporal demand item to avoid drawing participants' attention to task duration, which could have increased stress or anxiety. To enhance clarity and reduce visual complexity, we modified the original 21-point scale to a simplified 1–7 scale, removing the intermediate "low," "medium," and "high" labels associated with each point [21].

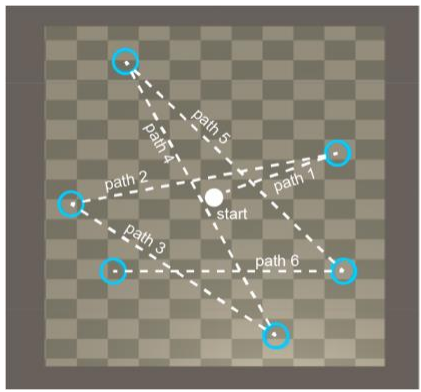

Figure 4. A bird's-eye view of the test environment is shown. Each participant began the task at the central "start" location on the checkered platform and then navigated to a series of targets, represented as blue circles. The six paths connecting the targets are labeled Path 1 through Path 6 to indicate the intended order of navigation. Image from Franz et al. (2023).

### 3.4   Design and Analysis

The study employed two, two-factor designs, which will be referred to as the *between-group comparison* and the *between-level comparison*. The first factor for both designs was *locomotion technique*. It was a within-subjects factor with six levels: Astral, Chicken, Grab, Sliding, Teleport, and Throw.

**Between-Group Comparison**: The second factor was *group* in this study design. It was a between-subjects factor with two levels: with impairment and without impairment.

**Between-Level Comparison:** The second factor was *impairment level* in this study design. It was a between-subjects factor with three levels: low, medium-low, and medium-high.

### 3.5   Performance Analysis

The three overall performance-related outcome measures were *trial time*, *hit rate*, and *obstacles hit*. Trial time represented the time to complete a trial, hit rate represented the number of targets participants hit successfully out of six, and obstacles hit referred to the number of obstacles a participant hit during the two trials in which obstacles were present. There were 8199 total data points in this analysis.

The responses to a post-task questionnaire were related to user experience and workload and were also analyzed as outcome measures. These included questions about *presence* and *simulator sickness*, as well as five items from the NASA-TLX workload questionnaire: *mental demand, physical demand, performance, effort,* and *frustration* [22].

There were 240 data points for all outcome measures except presence, which had 238 responses due to two participants not responding.

The aligned rank transform procedure for was used to analyze trial time [24] mixed-effects logistic regression was used for hit rate [17], mixed-effects negative binomial regression was used for obstacles hit [25] and ordinal logistical regression was used for post-task questionnaire outcomes [42]. Type I error rates were corrected using Holm's sequential Bonferroni procedure for *post hoc* comparisons [26].

**3.6 Metrics Analysis**

This section describes device-level data gathered throughout the study and introduces metrics aimed at identifying low-level differences in interactions between participants with and without impairments.

*3.6.1 Raw Data*

Device-level data were recorded from both motion controllers and the VR headset. The data included values positional, rotational, velocity, angular velocity, acceleration, and angular acceleration along the $x$, $y$, and $z$ axes for each device. Additionally, input data from controller buttons including thumbstick positions, trigger and grip presses, as well as primary and secondary button presses and touches were recorded. Positions of both controllers and the headset within the VE were also captured.

*3.6.2 Metrics: Overview*

We established several metrics related to movement, button interaction, and target acquisition by drawing upon prior research on measuring movement differences between individuals with and without impairments [12, 29, 33, 39, 44]. All metrics were designed to be agnostic to specific locomotion techniques to facilitate comparisons across different techniques and participant groups. We anticipated that certain metrics would only be relevant for specific techniques because of the techniques' distinct control methods. For example, metrics involving head movement were expected to apply primarily to techniques such as Chicken and Sliding. Therefore, differences observed between participant groups for particular techniques but not others would indicate that these differences arise from interactions specific to those techniques. A summary of all metrics examined is presented in Table 3.

Table 3. Each proposed metric, a description of the metric, its type, the reason the metric might capture differences between groups, and the locomotion technique for which the metric might show an effect. Note: "device" refers to the headset and right or left controller. Metrics are technique-agnostic.

| Metric | Type | Description | Rationale | Applicable Locomotion Technique |
|---|---|---|---|---|
| Device variability [33] | Movement-related | The total distance the device travelled | Could reflect the efficiency of movement | Chicken, Grab, Sliding, Teleport, Throw |
| Device extent [33] | Movement-related | The Euclidean distance between the two furthest device positions | Could reflect the range of motion and efficiency of movement | Chicken, Grab, Sliding, Teleport, Throw |
| Device velocity and acceleration | Movement-related | The average velocity and acceleration of the device | Could reflect the ability to control movement speed | Chicken, Grab, Teleport, Throw |
| Device angular velocity and acceleration | Movement-related | The average angular velocity and acceleration of the device | Could reflect the ability to control movement rotation speed | Chicken, Sliding |
| Total distance between device pairs [44] | Movement-related | Total Euclidean distance between device pairs (left-right, left-headset, right-headset) | Could reflect the efficiency of movement | Grab, Throw |
| Extent between device pairs [44] | Movement-related | Greatest Euclidean distance between device pairs | Could reflect range of motion | Grab, Throw |
| Number of submovements [29] | Movement-related | The number of movements between dips in velocity | Could reflect the efficiency of movement while performing the interaction | Chicken, Grab, Throw, |
| Thumbstick distance [33] | Button-related | The total distance the thumbstick traveled | Could reflect thumb strength and range of motion | Astral |
| Thumbstick extent [33] | Button-related | The angular distance between the two furthest angles | Could reflect thumb range of motion | Astral |
| Target re-entry [39] | Target-related | Number of times the user exited and re-entered the target | Could reflect the ability to control speed, direction, and position | All |
| Target axis crossings [39] | Target-related | Number of times the user crossed the optimal path to the target | Could reflect the ability to control speed, direction, and position | All |
| Movement variability [39] | Target-related | How straight the users' path is compared to the optimal path | Could reflect the ability to control speed, direction, and position | All |

*3.6.3  Movement-Related Metrics*

The metrics for *device variability* and *(angular) extent* were adapted from Kong et al.'s [33] methods for analyzing touch-based interactions. Device (angular) variability provides insights into how efficiently a participant executed movements, whereas device (angular) extent reflects both movement efficiency and the range of motion available in participants' torsos (via the headset) and arms (via the controllers). Participant interview responses from prior work [12] suggested that these metrics could highlight differences between groups. For example, users with impairments reported challenges turning their heads during Chicken and Sliding, which could be potentially captured in metrics like angular variability, extent, velocity, and acceleration measures for the headset. Additionally, participants noted difficulties reaching and repeatedly performing actions with Grab, which could have affected controller-based metrics like variability, extent, velocity, and acceleration. Device *angular velocity* and *acceleration* were calculated by averaging all data points within a trial. The other metrics were computed using the following formulas:

$$Device\ Variability = \sum_{i=1}^{n-1} \sqrt{(x_i - x_{i-1})^2 + (y_i - y_{i-1})^2 + (z_i - z_{i-1})^2} \in [0, \infty)$$

$$Device\ Extent = \max_{i,j \in 0,\dots,n-1} \sqrt{(x_j - x_i)^2 + (y_j - y_i)^2 + (z_j - z_{j-1})^2} \in [0, \infty)$$

Device angular extent was computed in a similar way, where R represents the Euler angle:

$$Device\ Angular\ Extent = \max_{i,j \in 0,\dots,n-1} \sqrt{(R_{jx} - R_{ix})^2 + (R_{jy} - R_{iy})^2 + (R_{jz} - R_{iz})^2} \in [0, 360]$$

The metrics for *device-pair variability* and *extent* were inspired by Pfeuffer et al.'s [44] research on behavioral biometrics in VR. In their study, the authors identified movement metrics that could distinguish between individuals' movement patterns for user authentication in VR. Their findings indicated that device-pair metrics were particularly effective as features for predictive models. Given the metrics' ability to differentiate between individual users, these metrics may also distinguish interactions between participants with and without impairments. The calculations for device-pair variability and extent are as follows:

$$Device\ Pair\ Variability =$$
$$\sum_{i=1}^{n-1} \sqrt{(x_{device1_i} - x_{device2_i})^2 + (y_{device1_i} - y_{device2_i})^2 + (z_{device1_i} - z_{device2_i})^2} \in [0, \infty)$$

$$Device\ Pair\ Extent =$$
$$\max_{i,j \in 0,\dots,n-1} \sqrt{(x_{device1_i} - x_{device2_i})^2 + (y_{device1_i} - y_{device2_i})^2 + (z_{device1_i} - z_{device2_i})^2} \in [0, \infty)$$

The *number of submovements* metric is based on Hwang et al.'s [29] work on analyzing movements while motion-impaired users interacted with a mouse. They defined a submovement as a movement within a rapid aimed movement where velocity dips before and after the movement (Figure 5). The number of submovements is the sum of submovements during a trial.

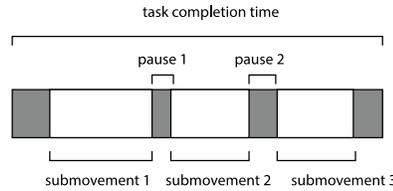

Figure 5. Submovements are movements within a rapid aimed movement with dips in velocity before and after their occurrence. Image adapted from [29].

The metric capturing the *number of submovements* was adapted from research by Hwang et al. [29], who studied cursor movement behaviors in users with motor impairments. They defined a submovement as a distinct segment within a quick aimed motion characterized by a reduction in velocity at both the start and end points (Figure 5). In the current study, the total number of submovements was calculated by counting all submovements observed throughout a single trial.

*3.6.4   Button-Related Metrics*

The button-related metrics were not derived from previous studies, but they potentially reflect various user abilities such as finger isolation, strength, control, and range of motion. Interviews revealed that some participants with impairments experienced difficulty turning their heads, leading them to rely heavily on thumbstick movements for turning, which could have been captured by these metrics. Additionally, participants mentioned challenges pressing buttons, including instances where fingers slipped off the buttons. The formulas for *thumbstick variability* and *extent* are:

$$Thumbstick\ Variability = \sum_{i=1}^{n-1} \sqrt{(x_i - x_{i-1})^2 + (y_i - y_{i-1})^2} \in [0, \infty)$$

$$Thumbstick\ Extent = \max_{i,j \in 0,\dots,n-1} \sqrt{(x_j - x_i)^2 + (y_j - y_i)^2} \in [0, 2]$$

Mean *grip* and *trigger pressure* values were calculated by averaging pressure data across each trial. *Button press counts* were found by totaling the number of presses for trigger, grip, primary, and secondary buttons throughout a trial.

*3.6.5   Target-Related Metrics*

Target-related metrics draw from MacKenzie et al.'s [39] metrics for assessing pointing-device accuracy but applied to the user's path in the VE, rather than a cursor's path. These metrics evaluate a user's ability to regulate their position, trajectory, and movement speed while navigating by comparing their actual path to an optimal straight-line path between start and end targets in the virtual test environment (Figure 6). Participants with impairments reported during interviews that they had poor control over their navigation with Chicken and Throw and struggled to judge their proximity to targets when using Chicken and Teleport. This feedback indicates that target-related metrics could help explain performance differences between users with and without impairments. Specifically, the first metric, *movement variability*, quantifies the deviation of participants' paths from the ideal straight-line path. Movement variability is calculated as:

$$Movement\ Variability = \sqrt{\frac{\sum_{i=1}^{n-1} y_i - \overline{y}}{n-1}}$$

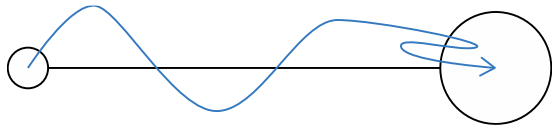

Figure 6. In this figure, which is a bird's-eye view of the target, the user's path is the blue line, and the optimal path is the horizontal black line. The starting point is on the left and the end target is on the right. The target re-entry count in this instance is 1 and the count of axis crossings is 2.

The second metric, *target re-entry count*, reflects the total number of times a participant left and re-entered the target area (Figure 6). The third metric, *axis crossings*, indicates how frequently a participant's path intersected with the optimal straight-line path (Figure 6).

*3.6.6 Design and Analysis*

The between-subjects factor, *group*, used in the metrics analysis was the same as the one in the performance evaluation. This factor consisted of two categories: participants with impairments and participants without impairments. The within-subjects factor, *technique*, included six categories representing each of the six locomotion techniques Two left-handed participants from the without impairments group were excluded from all analyses except those involving headset-related metrics due to them using the right instead of left controller. Additionally, because most participants with impairments did not use the left controller, we excluded these metrics from our analysis.

The metrics described previously served as outcome measures, with a separate calculation conducted for each metric per trial. This resulted in a total 2,721 data points for the right controller and 2,763 data points for the headset. An Anderson-Darling test [1] was conducted on the residuals of repeated-measures, full-factorial ANOVA models examining *group × technique*. Results indicated significant departures from normality for all metrics. Consequently, a nonparametric aligned rank transform approach [24, 57] with a linear mixed model [13] was used for the analysis. Type I error rates were corrected using Holm's sequential Bonferroni procedure for *post hoc* comparisons [26].

## 4 RESULTS

Results from overall performance and low-level metrics analyses are presented. Non-significant results are excluded, and omnibus test results can be found in Appendix A1 for the sake of brevity.

### 4.1 Performance Measures

There were some unexpected differences in overall performance between groups in terms of trial time, hit rate, and obstacles hit.

*4.1.1 Performance: Trial Time*

An Anderson-Darling test [1] was run on the residuals of repeated measures, full-factorial ANOVA models for *group × technique* and *impairment level × technique*. The tests were statistically significant ($A$=26.59, $p<.0001$; $A$=27.37, $p<.0001$, respectively), indicating that the residuals did not conform to a normal distribution. A nonparametric aligned rank transform procedure was used [24, 57] with a linear mixed model [13] for the analysis. *Post hoc* pairwise comparisons conducted with the ART-C procedure [7], indicated that all pairwise comparisons between groups were significantly different except for Sliding (Figure 7A). Post hoc comparisons between levels of impairment also revealed that only the Sliding technique showed no significant differences. This result suggests that participants with and without impairments reached targets within comparable times when using Sliding. For the remaining techniques, participants with medium-low and medium-high levels of impairment took significantly longer to navigate to targets than participants with low levels of impairment. "Medium-low" and "medium-high" participants had similar trial times for all techniques except Teleport, with which participants with medium-low levels of impairment had higher trial times than participants with medium-high levels of impairment (Figure 7B).

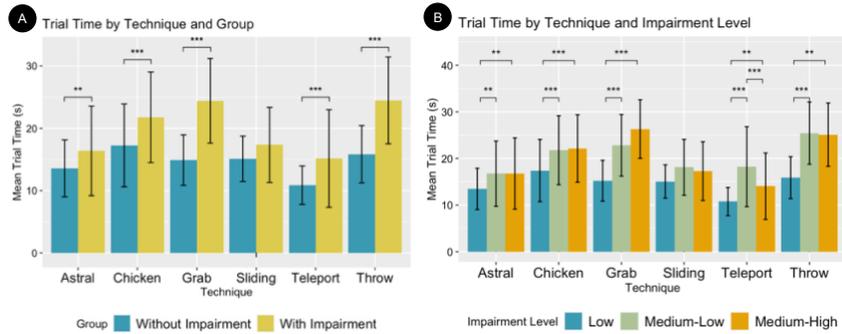

Figure 7. (A) Mean trial times were significantly higher for the group with impairments compared to the group without impairments for all techniques except Sliding. (B) Trial times differed significantly across impairment levels for all techniques except Sliding. Lower is better. For these and subsequent bar charts, significance codes are $p<.001$ (***), $p<.01$ (**), $p<.05$ (*).

*4.1.2 Performance: Hit Rate*

Because targets hit (versus missed) was a dichotomous dependent variable (i.e., it has a value of 0 or 1), we conducted an analysis using a mixed-effects logistic regression model [17, 51]. *Post hoc* pairwise comparisons indicated that participants with impairments had significantly lower hit rates than participants without impairments for all techniques except Sliding and Teleport (Figure 8A).

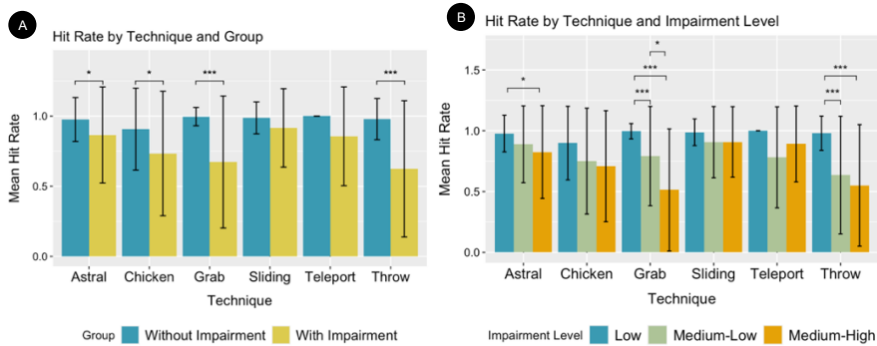

Figure 8. (A) Participants with impairments hit significantly fewer targets with all techniques except Sliding and Teleport. (B) Hit rates differed significantly across some impairment levels for all techniques except Chicken, Sliding and Teleport. Higher is better.

"Medium-high" participants hit significantly fewer targets than "low" participants for Astral, Grab, and Throw (Figure 8B). "Medium-low" participants hit significantly fewer targets for Grab and Throw compared to "low" participants, and "medium-high" participants hit significantly fewer techniques compared to "medium-low" participants for Grab.

*4.1.3    Performance: Obstacles Hit*

*Post hoc* pairwise comparisons indicated that the only significant difference in number of obstacles hit between groups was for Throw (Figure 9). One possible explanation for this result is that participants with impairments may have adopted a more cautious strategy, aiming for landing positions they could more reliably predict. As a result, they may have hit fewer targets than participants without impairments.

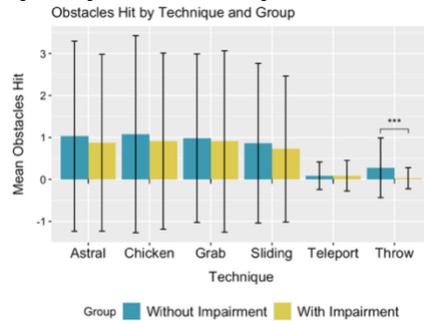

Figure 9. Participants without impairments hit significantly more obstacles with Throw compared to people with impairments. Lower is better.

**4.2    Post-Task Questionnaire**

In this section, we report the results of statistical tests comparing the *group* and *impairment-level* responses to the post-task questionnaire for each locomotion technique. All omnibus tests were conducted using an analysis of variance based on mixed-effects ordinal logistic regression [23] and *post hoc* pairwise comparisons for impairment level were performed using *Z*-tests, corrected with Holm's sequential Bonferroni procedure [26].

*4.2.1    NASA-TLX: Mental Demand*

*Post hoc* comparisons indicated that "medium-low" participants rated Teleport as being significantly more mentally demanding compared to "low" and "medium-high" participants (Figure 10).

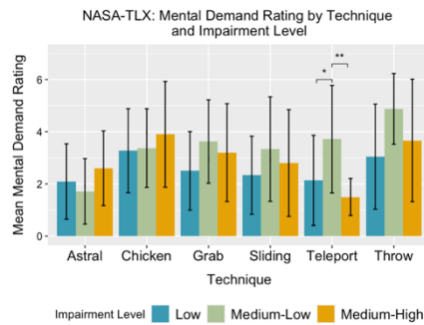

Figure 10. Mean mental demand rating (1-7) was higher for "medium-low" participants compared to the other groups for Teleport. Lower is better.

*4.2.2 NASA-TLX: Physical Demand*

There was a trending significant effect of *technique* and *group* on physical demand. Participants with impairments rated Astral, Grab, Teleport and Throw as being significantly more physically demanding than participants without impairments (Figure 11A). "Medium-high" participants thought Astral, Grab, and Throw were more physically demanding compared to "low" participants. "Medium-low" participants thought Teleport was more physically demanding than "low" participants and "medium-high" participants thought Grab was more physically demanding than "medium-low" participants. (Figure 11B).

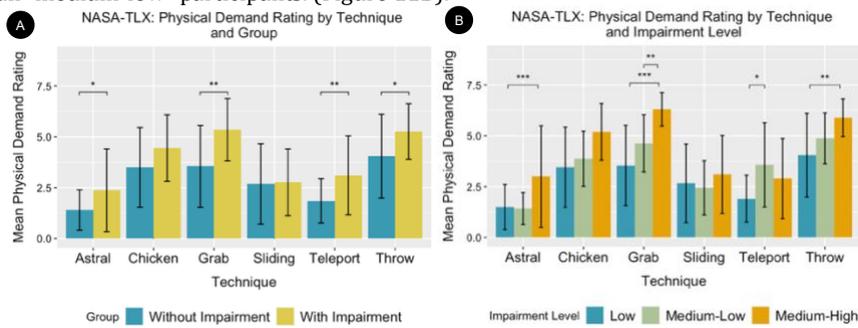

Figure 11. (A) Mean physical demand rating (1-7) was higher for people with impairments for Astral, Grab, Teleport, and Throw compared to people without impairments and (B) lowest for "low" participants compared to the other groups for the same techniques. Lower is better.

*4.2.3 NASA-TLX: Effort*

"Low" participants perceived Astral, Grab, and Teleport as requiring less effort to use than other groups (Figure 12).

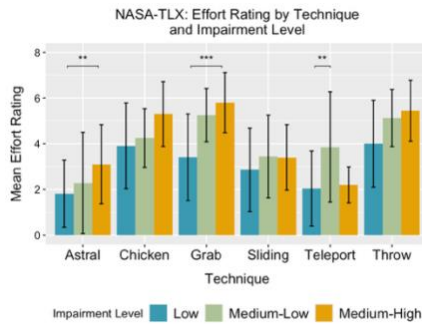

Figure 12. Mean effort rating (1-7) was lowest for "low" participants for Astral, Grab, and Teleport compared to other groups. Lower is better.

*4.2.4 NASA-TLX: Frustration*

People with impairments found Grab, Teleport, and Throw to be significantly more frustrating to use compared to people without impairments (Figure 13A). Comparisons also indicated that people with medium-high levels of impairment thought Grab was significantly more frustrating to use compared to people with low levels of impairment (Figure 13B).

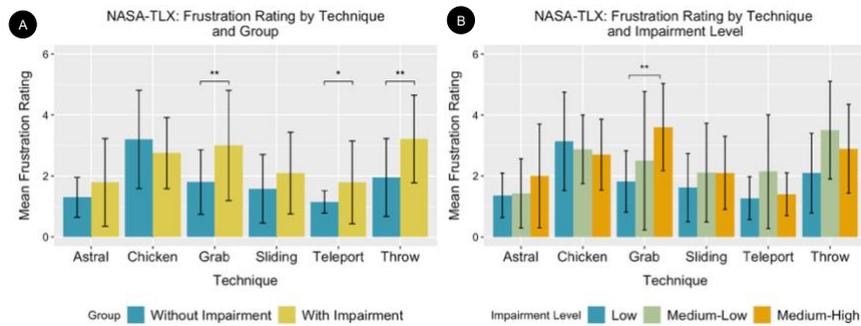

Figure 13. (A) Mean frustration rating (1-7) was higher for people with impairments for Grab, Teleport, and Throw and (B) highest for "medium-high" participants for Grab compared to the other groups. Lower is better.

*4.2.5   NASA-TLX: Performance*

"Low" participants thought they performed better with all techniques except Chicken compared to other groups (Figure 14).

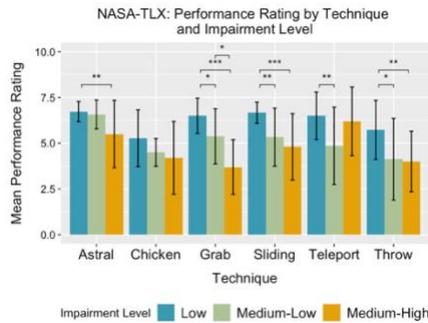

Figure 14. Mean performance rating (1-7) varied across impairment levels for all techniques except Chicken. Higher is better.

**4.3   Low-level Metrics Analysis**

The analysis of performance measures indicated that participants with impairments performed comparably to those without impairments when using Sliding and reported a similar perceived workload, suggesting that Sliding could be the most inclusive technique out of the six that we investigated. However, participants with impairments had lower performance and higher workload with the remaining techniques, suggesting that there are accessibility issues with these techniques that prevented people with impairments from performing similarly to the other group. Interview data from the prior study, which examined accessibility differences across six locomotion techniques among individuals with impairments, provided insights into how participants perceived their impairments influenced their interaction with VR. However, these perceptions were not supported by corresponding quantitative data [12]. The following analysis revealed some interesting differences in movement-, button-, and target-related metrics between groups that could explain differences in performance and lead to insights for designing more accessible locomotion techniques.

*4.3.1 Movement: Device Variability*

People with impairments had significantly higher head variability for all techniques except for Chicken, meaning that people with impairments moved their head a greater distance compared to people without impairments when using the other five techniques (Figure 15A), Chicken was the only technique with which participants leaned, suggesting that when they were intentionally moving their heads and torsos, the distances participants moved were similar between groups. Participants with impairments also had higher right controller variability for all techniques except Grab and Throw (Figure 15B). Grab and Throw were the only techniques with which participants had to extend their arms, also suggesting that when performing intentional movements, participants moved similar distances.

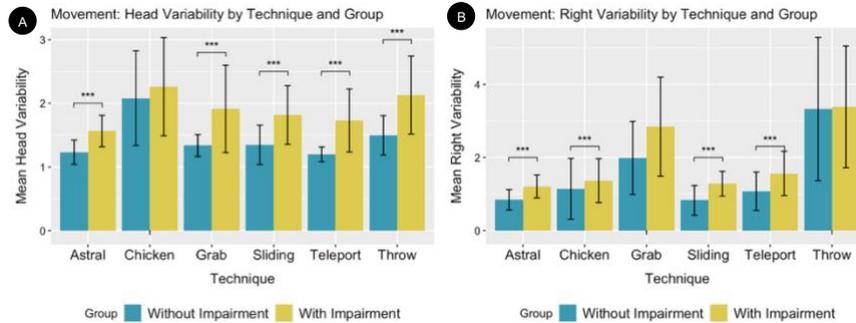

Figure 15. (A) Participants with impairments had significantly higher head variability for all techniques except Chicken, and (B) higher right controller variability for all techniques except Grab and Throw.

*4.3.2 Movement: Device Extent*

People with impairments had significantly higher head extent for all techniques (Figure 16A) and higher right controller extents for all techniques except Grab and Throw (Figure 16B). Right extent mirrored right variability, indicating that participants extended their arms to a similar maximum distance during intentional movements. In contrast, head extent did not align with head variability: participants with impairments generally exhibited a higher maximum head movement distance compared to those without impairments

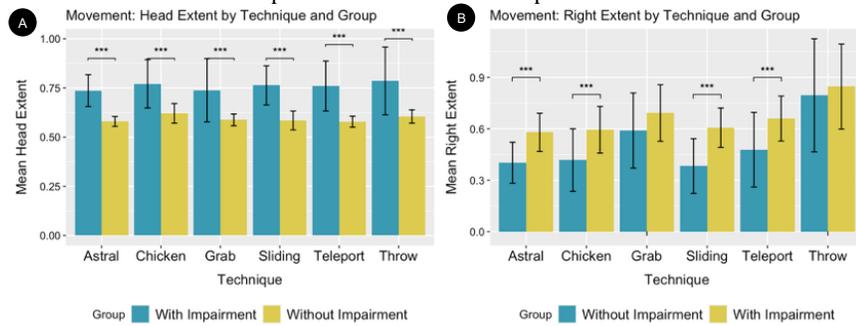

Figure 16. (A) Participants with impairments had significantly higher head extent values for all techniques and (B) higher right extent for all techniques except Grab and Throw.

*4.3.3   Movement: Device-Pair Variability*

Participants with impairments exhibited significantly greater head to right controller variability using Grab and Throw compared to those without impairments, indicating a consistently larger distance between the head and right controller throughout a trial. This finding could suggest that participants with impairments performed grab and throw gestures more frequently and at positions farther from their body center (Figure 17).

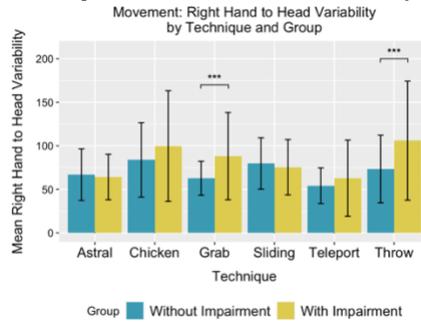

Figure 17. Participants with impairments had significantly head-right variability for Grab and Throw.

*4.3.4   Movement: Device Velocity and Accelerations*

Participants with impairments showed significantly lower head velocity during Chicken (Figure 18A) and lower head acceleration across all techniques (Figure 18B). These results suggest that, when using Chicken, they leaned their torsos forward more slowly than participants without impairments. Head acceleration might have been lower overall because people with impairments moved more gently and deliberately to achieve more stability while performing all techniques.

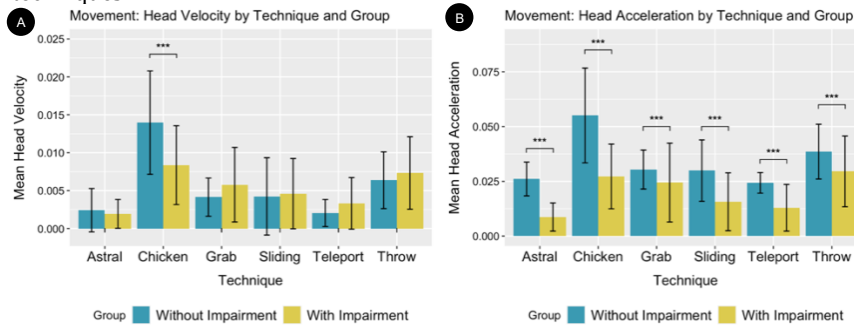

Figure 18. (A) Participants with impairments had significantly lower head velocity for Chicken and (B) lower head acceleration for all techniques.

*4.3.5   Movement: Angular Device Velocity and Acceleration*

People with impairments had significantly lower angular head velocity for Chicken and lower angular right velocity for Throw (Figure 19A). This finding indicates that participants rotated their heads more slowly compared to people without impairments when using Chicken, a technique in which movement direction is controlled by head rotation. Also, participants with impairments might have flicked their wrist more slowly when throwing the ball compared to people without impairments.

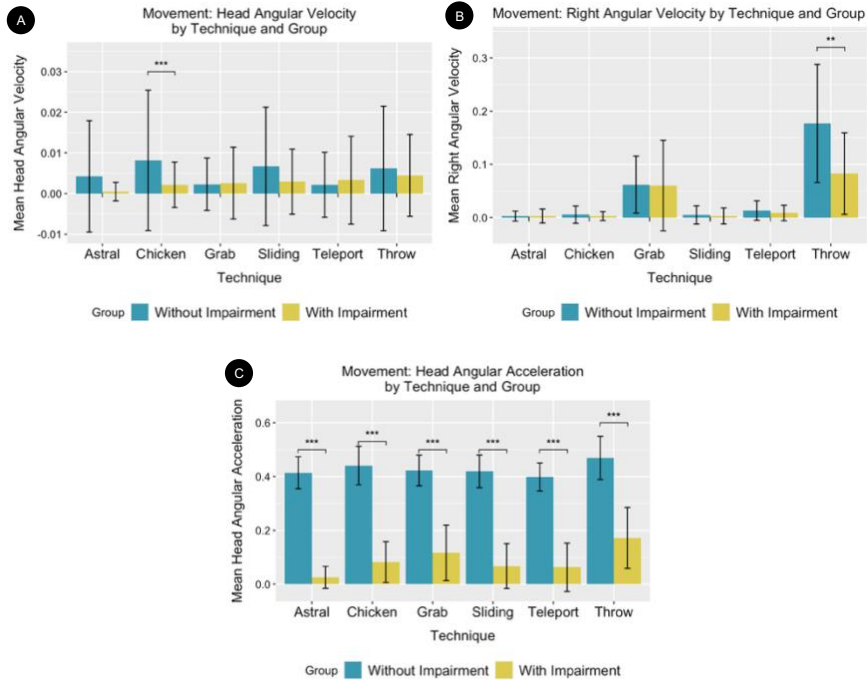

Figure 19. (A) Participants with impairments had significantly lower head angular velocity for Chicken (B), right angular velocity for Throw, and (C) lower head angular acceleration for all techniques

Participants with impairments had lower angular head acceleration for all techniques (Figure 19B). Like head acceleration suggests, significantly lower angular head acceleration could indicate that participants with impairments performed slower, more stable and deliberate movements with their heads and torsos when using all techniques.

*4.3.6   Movement: Submovement Count*

People with impairments had significantly more head submovements during Grab (Figure 20), indicating more frequent pauses between head movements. This finding suggests they may have made more frequent movement corrections or used more segmented motion when performing the grab and pull gesture.

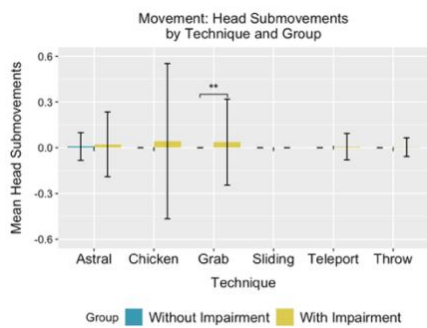

Figure 20. Participants with impairments had significantly more head submovements for Grab.

**4.4 Button-related Metrics**

This section presents the only significant button-related metric, thumbstick variability.

*4.4.1 Button: Thumbstick Variability*

People with impairments showed significantly higher right thumbstick variability during Teleport (Figure 21), indicating more frequent use of the thumbstick to rotate their view. Since the right thumbstick triggered snap turns, this suggests they may have relied more on rotational adjustments to correct their position or search for the target within the environment during a trial.

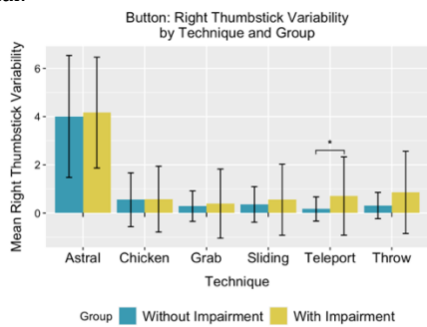

Figure 21. Participants with impairments had significantly higher right thumbstick variability for Teleport.

*4.4.2 Target: Target Re-Entry Count*

Participants with impairments had a higher target re-entry count for Astral and Grab (Figure 22), suggesting they had more difficulty maintaining control over their virtual position relative to the target compared to participants without impairments.

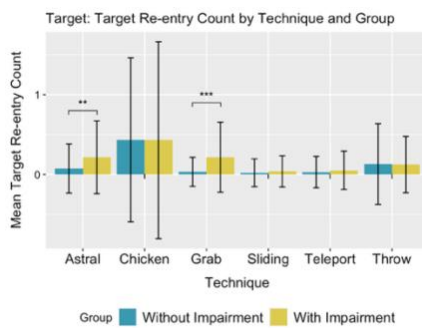

Figure 22. Participants with impairments had significantly higher target re-entry count for Astral and Grab.

## 5 DISCUSSION

Some head-related metrics revealed differences between groups regardless of technique. Other metrics can explain differences in performance for locomotion techniques.

### 5.1 Head-Related Metrics and Unintentional Movements Differentiate Groups

Higher head extent (Figure 16A), lower head acceleration (Figure 18B), and lower head angular acceleration (Figure 19C) were significantly different between groups across all techniques. These results suggest that participants with impairments exhibited gentler and more controlled, yet less centered, movements compared to those without impairments, indicating a more cautious but less centralized movement strategy. These three metrics could serve as indicators for distinguishing users with motor impairments, potentially enabling systems to recommend or automatically prompt the activation of accessibility features suited to their abilities (RQ.3).

A related finding is that participants with impairments appeared to make more unintentional movements. Participants with impairments moved their headset significantly a greater distance with all techniques except for the technique that was controlled exclusively with head movement, Chicken (Figure 15A).

This same pattern was apparent for right controller variability (Figure 15B). People with impairments moved their controllers a greater distance than people without impairments when they were using their controllers to perform techniques that did not require large arm movements (i.e., Astral, Chicken, Sliding, and Teleport). This finding is supported by prior research that found that older adults can overcome noise in their movements by applying greater force [55]. Also, people with intention tremor performed ballistic movements in a similar way to people without tremor [9]. These two findings suggest that noise in movement can be masked by intentional force, which could be an explanation for why intentional movements did not appear different in terms of device variability between groups in this study.

### 5.2 Need for Metrics to understand accessibility issues with VR interaction techniques

There were several surprising findings in this study. For example, the distance participants leaned forward over the course of a trial did not explain performance differences for Chicken. In a prior study [12], participants reported difficulty controlling their virtual navigation with Chicken, but the current results revealed that they were able to lean far enough, their performance differences were instead better explained by how quickly they leaned. Similarly, the total and maximum distance of arm movement did not account for performance differences between groups for the movement-based techniques Grab and Throw, despite participants previously citing limited shoulder range of motion and arm reach as challenges. However, a significant difference in head to right controller variability suggests that core muscle weakness could have affected some participants' ability to coordinate their arm and torso movements, also given the significant difference in head extent for these techniques (RQ.3). These findings highlight the value of using movement metrics to identify subtle differences between groups and to explain performance outcomes more precisely. While interview data provided a broad understanding of what might have made Chicken, Grab, and Throw less accessible, the quantitative metrics pinpointed the specific aspects of the interactions that were most challenging. These insights can be encouraging, as they suggest that participants' motor impairments did not prevent them from performing entire interactions, but rather affected certain components, which can be targeted and improved through design.

**5.3 Design Recommendations**

Based on our findings from analyzing performance, workload, and low-level metrics, we propose design recommendations. This section addresses RQ.3.

*Don't just make input accessible, make the virtual environment accessible too:* While prior work has primarily focused on adapting input interactions and hardware to make VR more accessible, our findings suggest that the design of the virtual environment itself can also play a critical role in accessibility. For example, with the Astral technique, participants with impairments had a significantly higher target re-entry count (Figure 22), indicating reduced control in keeping their avatars within the target area. One potential design solution to address this challenge is the introduction of *gravity wells*, which are subtle attractive forces centered on important objects or targets in the environment. That way, when avatars are close to the object, the user will need less manual control to move the avatar to the desired position as they will be "pulled" toward the object. Of course, gravity wells could have unwanted consequences as well.

With Teleport, Participants with impairments had greater right thumbstick variability than the other group (Figure 21). During the study, some participants overshot the target and then struggled to see where the target was when they were close to it. As a result, they had to re-orient their virtual bodies using the thumbstick to view the target. This metric reveals differences between groups in the corrective phase of the interaction. To support this phase, designers could implement features that help users locate nearby objects outside their field of view, such as directional arrows, audio cues, or subtle visual highlights.

The significantly greater number of head submovements and higher head-to-right-hand variability among participants with impairments suggest that head movements were more fragmented and that their right-hand actions were often decoupled from head movements during the Grab technique. This pattern indicates potential difficulty with coordinating upper body movements, which can make interactions more effortful and less fluid. Environmental modifications could enhance accessibility by helping to smooth out these interactions. For example, reducing ground friction in Grab and Pull could allow users to glide longer distances with less effort, thereby minimizing the need for repeated or tightly coordinated gestures.

*Adapt controls for timing, precision, and motor support:* Our findings align with prior work, such as Yamagami et al. [60], which emphasizes the value of providing computer assistance, such as mapping unimanual input to bimanual actions in VR, while also allowing users to control the level of automation. As an example, with Throw, participants without impairments had a greater right angular velocity (Figure 19B). One explanation is that they flicked their controller faster when releasing the ball than people with impairments during the throw motion, which could have impacted performance. During the study, several participants with impairments had difficulty timing the release of the grip button (which released the ball) with the end of their arm movement. Often, participants completed the throwing motion and then released the grip button afterwards, causing the ball to drop straight downwards. The accessibility of Throw could be improved predicting the best time to release the ball based on the user's arm trajectory and speed, so that a user does not need to time the release of the grip button at all.

In addition, participants with impairments' avatars had greater target re-entry count, suggesting less control over their avatars near the target. One potential design improvement would be to adaptively decrease joystick sensitivity as the avatar approaches a target, minimizing the likelihood that small joystick movements will cause the avatar to unintentionally exit the target area.

Furthermore, differences in head-related metrics between groups point to the possibility of detecting upper-body impairments through movement patterns. This finding suggests opportunities for systems to proactively recommend or enable accessibility features based on real-time user behavior.

*Allow for customization and minimal physical interaction:* Previous studies have emphasized the importance of designing gestures and interactions that require minimal physical effort and incorporate resting positions, as fatigue is a major barrier for people with motor impairments [53, 56, 59, 61]. Wentzel et al. [56] also highlight the value of allowing users to remap their physical range of motion to the virtual range required by the application. Our study supports these findings and further suggests that adjusting gain in movement-based locomotion techniques can significantly enhance accessibility. Specifically, by increasing the gain so that small physical movements result in larger virtual displacements, users with limited range of motion can navigate virtual environments more effectively and with less effort.

People with impairments had a lower head velocity and angular head velocity when using Chicken compared to participants with impairments (Figure 18A and Figure 19A), suggesting that the performance difference might have been related to participants leaning forward and rotating their heads more slowly, causing them to move more slowly in virtual space. Designers could make the Chicken technique more accessible by allowing users to adjust the gain for action, which would enable them to move and rotate further in the virtual environment with smaller physical movements (i.e., leaning forward less or turning their heads less) compared to the default settings. Similarly, participants with impairments exhibited a significantly greater distance between their headsets and right controllers when using Grab, suggesting that they may have reached forward more frequently than the other group. To accommodate this, designers could implement adjustable gain for the grab and pull interaction, allowing users to travel greater distances with shorter reaches.

*Don't assume bimanual control:* In line with Yamagami et al.'s [60] findings, we observed that despite the diversity of conditions among participants with impairments, most used only one controller during locomotion tasks. This highlights the importance of designing locomotion techniques that can be effectively operated with a single hand, while still supporting optional two-handed use, as seen with techniques like Grab and Throw. A key challenge with bimanual control is the risk of imbalanced movement. For example, with Grab, some participants who could not reach forward rotated their bodies to pull themselves sideways toward a target, using the direction their arm could extend. Similarly, with Throw, participants who could only throw to the side ended up moving laterally rather than forward. These cases underscore the need not only to support unimanual interaction, but also to consider how such interaction might result in asymmetrical or unintended movement patterns within the environment.

*Sliding: A Universally Inclusive Technique?:* The consistent trial times, hit rates, and obstacles hit across groups and impairment levels when participants used Sliding indicated no distinguishable performance differences, unlike the other five techniques (RQ.1). One possible explanation is that participants with impairments were more practiced in adapting to nonstandard interaction styles, such as using head movement to compensate for limited arm use, giving them an advantage in techniques like Sliding that allow for such adaptation. This finding suggests that Sliding may serve as a highly inclusive default locomotion technique, offering comparable performance across a range of physical abilities (RQ.2).

This finding also underscores the distinction between accessibility and inclusivity. In our prior work [12], Teleport emerged as the most accessible technique for participants with impairments; however, in this study, Sliding was the most inclusive. The difference stems from the fact that although participants with impairments performed well with Teleport, those without impairments performed even better, creating a performance gap. Furthermore, Teleport was perceived to have a significantly higher workload by participants with impairments and those with medium-low levels of impairment, which is in contrast to our prior study in which Teleport was reported as having the lowest workload overall. This result suggests that Teleport might be *relatively* easy for people with impairments, but disproportionately easier for those without impairments, making it less inclusive.

By contrast, Sliding not only produced similar objective performance across all groups but also resulted in similar perceived workload regardless of impairment severity. Interestingly, participants with medium-low and medium-high levels of impairment believed they performed worse than participants with low levels of impairment, despite no significant differences in actual performance. This discrepancy highlights that perceived performance is not always a reliable proxy for real performance. Ultimately, while accessibility refers to how well a system supports users within a specific group, inclusivity refers to how well a system equalizes performance across diverse user groups.

### 5.4 Lessons from the Study Design

In addition to study results, we also contribute some lessons we learned from designing a locomotion study in VR.

*Descriptiveness of trial time*: Trial time was the most important measure of performance for this experimental design. Trial time had a binomial distribution because of the 30 second time limit, and many participants failed to complete the trial. So, it can be seen as a finer-grained measure of hit rate. In other words, not only did trial time capture hit rate, but it also encoded how "close" the user was to a miss.

*Importance of Group vs. Impairment Level Comparisons*: Performance and workload differences often depended on impairment level, rather than the existence of an impairment. For example, there was not a significant difference in mental demand when looking at the group comparison for Teleport. But when looking at the impairment level comparison, there was a significant difference between medium-low and the other groups. This finding underscores the importance of treating functional impairment as a spectrum rather than a binary attribute in studies with people with disabilities.

*Reflecting on the Post-Task Questionnaire*: There was no effect of impairment on user experience-related measures of presence or sickness. Even though this is an intuitive finding, it actually contradicts previous work [20], which reported that presence might be influenced by having an impairment. People with multiple sclerosis (MS) experienced more presence when walking in a VE than did people without MS [20]. This contradiction could be a result of studying a more specific population (i.e., upper-body impairment vs. MS) or the locomotion technique used (i.e., seated locomotion techniques vs. real walking). Future work can explore how different types of impairment and locomotion impact presence.

### 5.5 Limitations and Future Work

While these findings offer valuable insights into how movement patterns can signal motor impairments and inform adaptive accessibility features, they also raise significant privacy concerns. The ability to infer sensitive health information, such as the presence of motor impairments, based solely on head and controller movement data presents a risk of unintended biometric profiling. If such data are collected without informed consent or used beyond their intended scope, users could be identified, categorized, or discriminated against based on physical ability without their knowledge.

Future analysis could also investigate how low-level metrics vary across impairment levels based on *Quick*DASH scores, where *impairment level* is used as an independent variable. The current analysis did not require a finer grained analysis of *impairment level* to reveal interesting results, but could be relevant to other studies, for example, to investigate how much a metric is able to distinguish levels of impairment.

## 6 CONCLUSION

People with impairments had similar performance and workload compared to people without impairments when using Sliding, where the only difference found between impairment levels was *perceived*—not actual—performance. This finding suggests that Sliding could be used as a default technique in VR applications.

Some head-related metrics revealed differences between participant groups across all techniques, suggesting that these metrics might be used to identify whether a user has an impairment. Differences in performance with locomotion techniques could be explained by some movement-, button-, and target-related metrics, providing insight into why participants with impairments performed differently compared to those without impairments. Understanding how and why people with and without impairments perform differently with locomotion techniques can inform the design of more accessible virtual reality experiences, ensuring that everyone has equal opportunities to benefit from emerging technologies like VR.

## 7 ACKNOWLEDGMENTS

This work was supported by a Facebook Social VR grant and by the Apple AI/ML Ph.D. Fellowship. Any opinions, findings, conclusions, or recommendations expressed in our work are those of the authors and do not necessarily reflect those of any supporter.

# A APPENDICES

## A.1 *Quick*DASH Questionnaire

### QuickDASH

Please rate your ability to do the following activities in the last week by circling the number below the appropriate response.

| | NO DIFFICULTY | MILD DIFFICULTY | MODERATE DIFFICULTY | SEVERE DIFFICULTY | UNABLE |
|---|---|---|---|---|---|
| 1. Open a tight or new jar. | 1 | 2 | 3 | 4 | 5 |
| 2. Do heavy household chores (e.g., wash walls, floors). | 1 | 2 | 3 | 4 | 5 |
| 3. Carry a shopping bag or briefcase. | 1 | 2 | 3 | 4 | 5 |
| 4. Wash your back. | 1 | 2 | 3 | 4 | 5 |
| 5. Use a knife to cut food. | 1 | 2 | 3 | 4 | 5 |
| 6. Recreational activities in which you take some force or impact through your arm, shoulder or hand (e.g., golf, hammering, tennis, etc.). | 1 | 2 | 3 | 4 | 5 |

| | NOT AT ALL | SLIGHTLY | MODERATELY | QUITE A BIT | EXTREMELY |
|---|---|---|---|---|---|
| 7. During the past week, *to what extent* has your arm, shoulder or hand problem interfered with your normal social activities with family, friends, neighbours or groups? | 1 | 2 | 3 | 4 | 5 |

| | NOT LIMITED AT ALL | SLIGHTLY LIMITED | MODERATELY LIMITED | VERY LIMITED | UNABLE |
|---|---|---|---|---|---|
| 8. During the past week, were you limited in your work or other regular daily activities as a result of your arm, shoulder or hand problem? | 1 | 2 | 3 | 4 | 5 |

Please rate the severity of the following symptoms in the last week. *(circle number)*

| | NONE | MILD | MODERATE | SEVERE | EXTREME |
|---|---|---|---|---|---|
| 9. Arm, shoulder or hand pain. | 1 | 2 | 3 | 4 | 5 |
| 10. Tingling (pins and needles) in your arm, shoulder or hand. | 1 | 2 | 3 | 4 | 5 |

| | NO DIFFICULTY | MILD DIFFICULTY | MODERATE DIFFICULTY | SEVERE DIFFICULTY | SO MUCH DIFFICULTY THAT I CAN'T SLEEP |
|---|---|---|---|---|---|
| 11. During the past week, how much difficulty have you had sleeping because of the pain in your arm, shoulder or hand? *(circle number)* | 1 | 2 | 3 | 4 | 5 |

*Quick*DASH DISABILITY/SYMPTOM SCORE = $\left(\left[\dfrac{\text{(sum of n responses)}}{n}\right] - 1\right) \times 25$, where n is equal to the number of completed responses.

A *Quick*DASH score may **not** be calculated if there is greater than 1 missing item.

Figure 23. The *Quick*DASH questionnaire used to measure upper-body impairment. *Quick*DASH is a validated questionnaire [19].

## A.2 Technology-Related (Physical) Impairment Questionnaire (TRIQ)

Check all the impairments that you experience:

- Slow movements
- Low strength in core, shoulders, neck, arms, hands or fingers
- Tremor
- Poor coordination
- Rapid fatigue when using core, shoulders, neck, arms, hands or fingers
- Difficulty gripping objects
- Difficulty holding objects
- Lack of sensation in core, shoulders, neck, arms, hands or fingers

- Difficult controlling direction of hand or arm movement
- Difficulty controlling distance of hand or arm movement
- Limited range of motion in core, shoulders, neck, arms, hands, wrists or fingers
- Pain in core, shoulders, neck, arms, hands, wrists or fingers
- Poor precision with finger(s)
- Poor finger isolation
- Limited wrist extension or flexion
- Difficulty moving core, shoulders, neck, arms, hands or fingers quickly
- Difficulty moving core, shoulders, neck, arms, hands or fingers at the right time

- Difficulty balancing while seated
- Limited mobility of lower body (legs or feet)

Figure 24. The **T**echnology-**R**elated Physical **I**mpairment **Q**uestionnaire (TRIQ) questionnaire used in the locomotion studies. It is not validated but based on literature in accessible computing [8, 43].

## A.3 Summary of Statistical Test Results for Dependent Variables

Table 4. Statistical test results for performance and post-task questionnaire measures by group (with impairment, without impairment) and impairment level (low, medium-low, medium-high).

| Measure | Group | Impairment Level |
|---|---|---|
| Trial Time | $F(5, 63.7) = 161.58, p<.001$ | $F(10, 2679.9) = 39.56, p<.001$ |
| Hit Rate | $\chi^2(5, N=2733) = 26.66, p<.001$ | $\chi^2(10, N=2733) = 58.55, p<.001$ |
| Obstacles Hit | $\chi^2 (5, N=2733) = 22.73, p<.001$ | n.s. |
| Presence | n.s. | n.s. |
| Sickness | n.s. | n.s. |
| Mental Demand | n.s. | $\chi^2 (10, N=237) = 29.02, p<.01$ |
| Physical Demand | $\chi^2 (5, N=237) = 10.88, p = .053$ | $\chi^2 (10, N=237) = 25.12, p<.01$ |
| Effort | n.s. | $\chi^2 (10, N=237) = 19.30, p<.05$ |
| Frustration | $\chi^2 (5, N=237) = 16.17, p<.01$ | $\chi^2 (10, N=237) = 23.34, p<.01$ |
| Performance | n.s. | $\chi^2 (10, N=237) = 25.64, p<.01$ |

Table 5. Statistical test results for movement metrics by device (head, right controller).

| Metric | Head | Right Controller |
|---|---|---|
| Device Variability | $F(5, 2780.2) = 29.99, p<.001$ | $F(5, 2786.2) = 11.87, p<.001$ |
| Device Extent | $F(5, 2778.3) = 15.56, p<.001$ | $F(5, 2785.7) = 67.79, p<.001$ |
| Device Velocity | $F(5, 2779.9) = 60.25, p<.001$ | n.s. |
| Device Acceleration | $F(5, 2779.9) = 49.44, p<.001$ | n.s. |
| Device Angular Velocity | $F(5, 2781.4) = 146.99, p<.001$ | $F(5, 2781.4) = 146.99, p<.001$ |
| Device Angular Acceleration | $F(5, 2780.1) = 32.41, p<.001$ | n.s. |
| Device Submovements | $F(5, 2983.33) = 288.77, p<.001$ | n.s. |

Table 6. Statistical test results for movement, button and target metrics

| Metric | Result |
|---|---|
| Head-Right Variability | $F(5, 2785.9) = 40.21, p<.001$ |
| Right Thumbstick Variability | $F(5, 2787.7) = 11.26, p<.001$ |
| Right Thumbstick Extent | n.s. |
| Target Re-Entry Count | $F(5, 2791.1) = 5.44, p<.001$ |
| Target Axis Crossings | n.s. |
| Movement Variability | n.s. |